\documentclass[11pt,conference]{IEEEtran}
\makeatletter
\def\ps@headings{%
\def\@oddhead{\mbox{}\scriptsize\rightmark \hfil \thepage}%
\def\@evenhead{\scriptsize\thepage \hfil \leftmark\mbox{}}%
\def\@oddfoot{}%
\def\@evenfoot{}}
\makeatother
\usepackage{cite}
\usepackage[dvips]{graphicx}
\usepackage[cmex10]{amsmath}

\newtheorem{theorem}{Theorem}[section]
\newcommand{\bTheorem}{ \begin{theorem}  }
\newcommand{\eTheorem}{ \end{theorem}    }

\newcommand{\bProof}{ \noindent {\bf Proof:} }
\newcommand{\eProof}{\hspace*{.1in} \hfill \begin{picture}(6,6)
\thicklines \put(0,0){\line(0,7){7}} \put(1,0){\line(0,7){7}}
\put(1.5,0){\line(0,7){7}} \put(2,0){\line(0,7){7}}
\put(3,0){\line(0,7){7}} \put(4.5,0){\line(0,7){7}}
\put(4,0){\line(0,7){7}} \put(5,0){\line(0,7){7}}
\end{picture} }

\def\BibTeX{{\rmfamily B\kern-.05em{\scshape i\kern-.025em b}\kern-.08em \TeX}}

\newcommand{\bEq}{ \begin{eqnarray}  }
\newcommand{\eEq}{ \end{eqnarray}    }

\newtheorem{proposition}{Proposition}[section]
\newcommand{\bProposition}{ \begin{proposition}  }
\newcommand{\eProposition}{ \end{proposition}    }

\newtheorem{Definition}{Definition}[section]
\newcommand{\bDefinition}{ \begin{Definition} }
\newcommand{\eDefinition}{ \end{Definition} }
\newcommand{\bDef}{ \begin{Definition} }
\newcommand{\eDef}{ \end{Definition} }

\newtheorem{lemma}{Lemma}[section]
\newcommand{\bLemma}{ \begin{lemma}  }
\newcommand{\eLemma}{ \end{lemma}    }

\newtheorem{Remark}{Remark}[section]
\newcommand{\bRemark}{ \begin{Remark} \rm }
\newcommand{\eRemark}{ \end{Remark}    }

\newtheorem{corollary}[theorem]{Corollary}

\begin{document}
%
\title{A Stable Fountain Code Mechanism for Peer-to-Peer Content Distribution}

\author{\IEEEauthorblockN{Cedric Westphal}
\IEEEauthorblockA{Innovation Center, Huawei Technologies\\Santa Clara, CA\\
Computer Engineering Department, University of California\\Santa Cruz, CA\\
Email: cedric.westphal@huawei.com, cedric@soe.ucsc.edu}
}

\maketitle

\begin{abstract}
Most peer-to-peer content distribution systems require the peers to privilege the welfare of the overall system over greedily maximizing their own utility. When downloading a file broken up into multiple pieces, peers are often asked to pass on some possible download opportunities of common pieces in order to favor rare pieces. This is to avoid the missing piece syndrome, which throttles the download rate of the peer-to-peer system to that of downloading the file straight from the server. In other situations, peers are asked to stay in the system even though they have collected all the file's pieces and have an incentive to leave right away.

We propose a mechanism which allows peers to act greedily and yet stabilizes the peer-to-peer content sharing system. Our mechanism combines a fountain code at the server to generate innovative new pieces, and a prioritization for the server to deliver pieces only to new peers. While by itself, neither the fountain code nor the prioritization of new peers alone stabilizes the system, we demonstrate that their combination does, through both analytical and numerical evaluation.
\end{abstract}

\section{Introduction}

New network architectures are being studied  to facilitate content distribution\cite{Jacobson2009Networking,Zhang2010Named,Pursuit,Ahlgren2012Survey,azimdoost2012throughput,chanda2013content,chanda2013contentflow,su2013benefit}. In many of these proposals, the mechanisms to implement content distribution propose to take advantage of cached copies of the content located in distributed locations in the network, including end users and other peers. Some CDNs\cite{Akamai2007} even have started using P2P distribution systems in order to reduce downloading costs. Those new usages of P2P systems would come on top of the existing popularity of P2P systems such as BitTorrent\cite{BitTorrent} or eMule\cite{eMule}, which account for a significant fraction of the overall Internet traffic already.

The point of using peers to share content is that the multiplicity of the potential peers would allow the content distribution to scale: the more peers, the higher the rate of downloads. A server offers a file to download, which is divided in a number of chunks. Peers (or leechers) join in the system, request chunks from both the server and other peers. Then, once a peer has obtained all the chunks from the requested file, it leaves the system.

However, for some peer-to-peer systems, the expected benefit from the participation of peers turns out to be elusive\cite{Hajek2010Missing}\cite{Mathieu2006Missing}. The so-called {\em missing chunk syndrome} ends up restricting the download rate to that of the server. In this case, there is no gain from the peer diversity, since the rate of download is the same as if all peers were pulling the file from the server.

Some solutions to alleviate this problem have been recently proposed\cite{Oguz:EECS-2012-6}\cite{Zhu2011Stability}\cite{Zhou2011Stability}. These solutions work by either restricting which chunk a peer can download (say, constraining the peer to download only rare pieces first) or forcing the peers to stay longer in the system (as a seeder). Either way, the peer ends being coerced into a behavior that is good for the system as a whole, but not necessarily for each of the individuals.

We propose a mechanism which lets the peer function in a greedy manner, namely download any chunk that it does not have upon each contact with another peer; and leave the system as soon as it has received the whole target file. We also consider a pure random peer selection, meaning that peers do not need to poll multiple neighbors, nor need to maintain some chunk lists for their neighbors. In essence, we consider the simplest selection mechanism from the peer's point of view. Our method combines two improvements, which do not work independently of each other. Namely, we suggest to use a fountain code\cite{Byers:1998} at the server to insert chunks with more degrees of freedom, and to have the server push chunks preferentially to new peers, when the rate of arrivals of new peers into the system exceeds the server's download capacity.

Fountain codes (or rateless erasure codes) take the file to be downloaded as input and produce a sequence of chunks such that obtaining any $k$ elements of the sequence allows to reconstruct the original file. The use of coding in content distribution networks\cite{Gkantsidis2005Network} and information-centric networks\cite{Montpetit2011Network} has demonstrated positive benefits, and our peer-to-peer system proves to be no exception.

We demonstrate that the simple combination of coding and prioritizing of the server's download does indeed increase the capacity of the peer-to-peer system while imposing no restrictions on the behavior of the peers. We show that by proper dimensioning of its parameters, any peer workload can be stabilized. While our result is primarily theoretical, we confirm these findings by simulating our proposed mechanism and verify that it does increase the system's download capacity.

The paper is organized as follows: in Section~\ref{sec:rel}, we discuss the related work and offer some background. In Section~\ref{sec:model}, we describe the model of our peer-to-peer system and state our main theorem. In Section~\ref{sec:proof}, we offer the proof of the Theorem. We visualize these results using Matlab in Section~\ref{sec:num} and finally provide some concluding remarks in Section~\ref{sec:con}.

\section{Related Work}
\label{sec:rel}

Consider a system where the server does not use any coding and delivers one of the $k$ file pieces to the peers. More precisely, the peers contact at random another peer, including the server, and download a chunk if the other peer has an innovative piece. The peers leave the system as soon as they complete the download. This will serve as our {\em baseline system}, which we will use as a benchmark.

The missing chunk syndrome\cite{Hajek2010Missing}\cite{Mathieu2006Missing} arises when a disproportionate number of peers in the system have all the same chunks but one. These nodes are denoted as the `one-club' peers.

\begin{figure}[!t]
\centering
\includegraphics[width=3in]{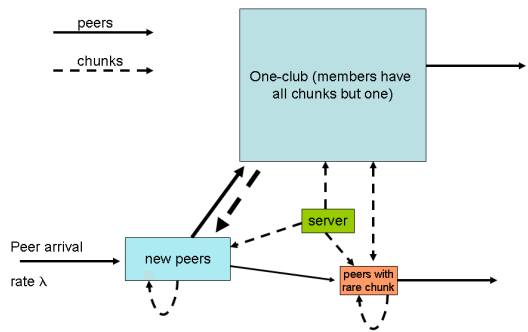}
\caption{The Missing Chunk Syndrome (reproduced from\cite{Mathieu2006Missing})}
\label{fig:missing}
\end{figure}

In this case, the last chunk is a rare chunk, and nodes with a rare chunk will leave the system fast, as they can get all the other pieces from the `one-club' peers. Thus, nodes with the rare chunk will depart from the system, leaving only the server to provide the rare chunk. And each `one-club' peer that receive the "missing chunk" will depart the system right away since it will have completed the download of the whole file. As a consequence, the download performance of the system boils down to that of the server.

BitTorrent alleviates this issue by asking peers to poll a set of neighbors for their chunk list, and download the rarest chunk. This adds an overhead to share the availability of chunks to the peers in the neighbor set. Further, Oguz\cite{Oguz:EECS-2012-6} mentions that it is still an open question to know if BitTorrent protocol is stabilizing.

\cite{Oguz:EECS-2012-6} does provide a stabilizing protocol which requires to poll only three peers. It works by asking peers with no pieces to hold on downloading a piece until they find a rare piece, where a rare piece is defined by sampling three neighbors and if a chunk appears only in one of the three neighbors' chunk lists, then it is rare. A similar rule applies to peers missing only one chunk.

\cite{Zhu2011Stability} proves the stability of the P2P system when the peers are asked to stay in the system for some time after download completion that depends on the peer arrival process and download rate. As in the previous system, peers are asked to stay longer, either explicitly or by withholding download, in order to stabilize the system.

\cite{Qiu2004Modeling} models a BitTorrent P2P network and studies its scalability using a fluid model. It also models the peer selection mechanism and shows the convergence of the peer selection mechanism to a Nash equilibrium under some incentive structure. The paper only analyzes the BitTorrent protocol, and does not propose a novel mechanism. The fluid model was also studied in~\cite{Chow2009}.

Yang\cite{Yang2004Service}\cite{Yang2006} also analyzes the service capacity of peer to peer networks in both a transient and a stationary regime, and demonstrates that in both regimes the system is scalable. However, it does not consider the missing chunk syndrome but only an aggregate capacity over multiple files.

\cite{Massoulie2005Coupon} generalizes the peer-to-peer setup to coupon collection, and studies the asymptotic behavior of such systems with respect to the sojourn time in the system under different types of encounters, including random encounters (which is what we consider in this paper). This work also considers the missing chunk syndrome but under a closed system.

\cite{Norros2011Stability} studies the stability in the case of two-chunks systems. This is complementary to our works, as the benefit of our approach depend on having a file decomposed into a large number of chunks (namely, as the available bandwidth scales with the number of chunks $k$ to a file).

\cite{Menasche2011Implications} studies the impact of peer-selection and piece-selection policies. Sophisticated peer- and/or piece-selection requires the publisher of the file and/or the peers to maintain track of the different chunks in the swarm, in order to enable most-deprived peer selection and rarest-first piece selection. These selection mechanisms improve the performance of the system, but we consider an orthogonal problem, namely we impose a random piece- and peer- selection.

\cite{Zhou2011Stability} observes that missing chunk syndrome leads to a  bandwidth bottleneck at the seed that can lead to the underutilization of the aggregate capacity, and proposes to share this capacity across the download of multiple files.

\cite{baccelli2013p2p} studies the stability of P2P system, but taking into account the locality of the peers and the RTT between them to observe that P2P system might exhibit what they denote as super-scalability, namely a reduction of the delays as the number of peers grow.

We describe our model in more details in the following Section.

\section{Model}
\label{sec:model}

We consider the following system. Let $S$ be the total number of peers in the system. Peers join the system according to a process with intensity $\lambda$. We make the following assumptions: We consider slotted arrivals, with on average $\lambda$ arrivals per slot. We also assume that the number of arrivals in a time slot is greater than 1 and less than some $A>0$. These are technical assumptions which can be relaxed, but makes our proof simpler to explain. Peers come into the system empty. There is only one file in the system to download (multiple files can be considered independently). The server generates chunks from the file using a fountain code such that retrieving any $k$ chunks allows to decode the file. With a perfect fountain code, the server could generate a limitless sequence of such chunks. In our evaluation, we assume that the server creates $K>>k$ possible chunks (say, using linear random network coding over a large enough Galois field)\footnote{We ignore here the issues of efficient decoding inherent to using a true fountain code vs a random linear code, as those are orthogonal to the stability aspects we investigate.}.

The peers exchange chunks according to the following mechanism. It is a slotted system, where a slot corresponds to the exchange of one chunk. During each slot, a peer $p$ selects at random a peer $q$ among all the other peers in the system. If $q$ has one or more pieces of the file that $p$ does not have, then $p$ downloads one of these pieces, selecting at random if there are several matches. The download occurs immediately. If $q$ has no innovative element for $p$, then no download occurs.

When $p$ has collected $k$ distinct pieces, then it can decode the target file, and it leaves the system right away. According to this policy, every time $p$ can download any chunk, it does so; and it leaves the system at the first opportunity. The mechanism is myopic for $p$, as $p$ only maximizes its own utility and does not worry about the overall system's welfare.

\begin{figure}[!t]
\centering
\includegraphics[width=3in]{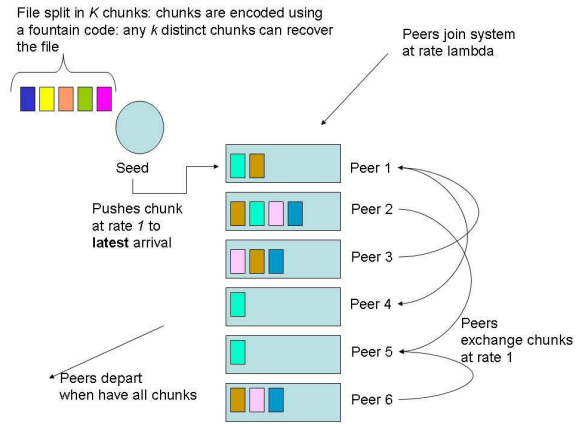}
\caption{P2P System Description}
\label{fig:p2pfountain}
\end{figure}

The server follows a different policy: when a new peer joins the system, it pushes a new chunk to this peer. However, the server is bandwidth constrained, and can only do this up to a rate of 1. This means, when $\lambda > 1$, we decompose the peer arrival process into two streams: one of rate 1 and one of rate $\lambda -1$ (this is achieved by tossing a coin selecting the first stream with probability $1/\lambda$ for the first one, and $(\lambda-1)/\lambda$ for the second one). The elements in the former process receive an innovative chunk from the server, while the latter ones join the system empty and follow the peer process as described above\footnote{Even though there is a restriction on which peers the server can serve, there is no "withholding download" since for an arrival rate of $\lambda > 1$, there is almost always a new peer for the server to serve. If there is no such new peer, the server can always pick another peer at random.}. We only consider $\lambda > 1$ as otherwise, the stability of the system is known from\cite{Zhu2011Stability}.

Each peer downloads a chunk at rate 1 if it finds a valid piece. On average, each peer is probed at rate one, so both upload and download bandwidth at each peer are identical and equal to one on average.

Our system describes a Markov process with state defined by the peer chunk profiles. Using the typical notation, denote by $q(x,x')$ the generator of this Markov process. As in~\cite{Zhu2011Stability}\cite{Oguz:EECS-2012-6}, we will use the Foster-Lyapunov criterion to show that the Markov process is positive recurrent.

Recall the following definition
\cite{Foss2008Lyapunov}. Let $V:S \rightarrow \mathbf{R}_+$ be some function on the state space $S$ of the Markov chain:
\bDef
The drift $\Delta V(x)$ of $V(x)$ is defined as:
\begin{eqnarray} \Delta V(x) = \Sigma_{x' \neq x} q(x,x')(V(x')-V(x))
\end{eqnarray}
\eDef

We enounce the following Theorem (Foster-Lyapunov)\cite{Foss2008Lyapunov}:
\bTheorem
\label{theo:drift}
Let $L$ be a function on the state space with drift $\Delta L$ with $L \geq 0$ and $\{L \leq l\}$ finite for any $l>0$. If there exists $S_o > 0$ and $\epsilon > 0$ such that, for $S>S_o$, $\Delta L < - \epsilon$, then the underlying Markov process is positive recurrent.
\eTheorem

We can now state our main result:
\bTheorem
\label{theo:stable}
For $\lambda < k$, the peer-to-peer model described above is stable
\eTheorem

Before proving the theorem in the next Section, we make a few observations, namely that neither the fountain code nor the prioritization independently increase the stability region of the baseline system. Both need to be combined to achieve the increased stability region. We describe this intuitively.

{\bf The "server prioritizes peers with no chunks" policy is not enough:} Consider again the baseline system where $k$ chunks are required to complete the download. Assume now that the server only serves nodes with no chunks, while the peers proceed according to the baseline mechanism. Again, when a `one-club' arises (it will happen at some point as the Markov process is irreducible and there is a positive probability of reaching an unbalanced state), all  nodes are missing one element. The server will provide this element at rate $1/k$ as it provides all $k$ elements at rate 1. Nodes which receives no chunk or a chunk that is not rare from the server will eventually join the `one-club' by downloading pieces from this predominant group.

 After a node with no chunk receives the rare element from the server, it will download the other elements. Since most of the peers belong to the `one-club', the node with the rare chunk will acquire the $k-1$ elements from the `one-club'. The rate at which nodes in the `one-club' acquire the rare chunk from peers with the rare chunk and $i-1$ other chunks is $c_{r,i} * C_{oc}/S$ where $S$ is the total number of peers, $c_{r,i}$ is the number of nodes with the rare chunk and $i-1$ other chunks, for a total of $i$ chunks, and $C_{oc}$ is the number of nodes in the `one-club'. This is because $C_{oc}$ nodes select a node with the rare chunk and $i$ pieces with probability $c_{r,i}/S$. However, this rate also approximates the rate at which peers with the rare chunk go from $i$ chunks to $i+1$ by downloading a piece from the `one-club'.

Because nodes with the rare chunks never join the `one-club' by definition, nodes inserted by the server with the rare chunk at rate $1/k$ will transition from $i$ chunks to $i+1$ chunks at the same rate. The rate at which these rare nodes transition from $i$ to $i+1$ chunks is equal to the rate at which they "knock down" peers to leave the `one-club' and thus the system. As there are $k$ steps for a rare peer to go from 1 chunk to $k$ chunks, it will reduce the size of the `one-club' by $k * 1/k = 1$, that is the server capacity. It is thus equivalent for the `one-club' to download straight from the server as in the baseline system.

{\bf The fountain code by itself is not enough:} Consider one more time the baseline system, but now, the server always provides innovative packets under some fountain code. Again, this is a Markov process that is irreducible, and there is a positive probability of reaching a `one-club' state, namely a state where all the peers have the same identical $k-1$ packets. These peers cannot find new chunks in between themselves, so only peers connecting to the server will acquire a $k$-th packet, which allows them to decode the file with the already obtained $k-1$ pieces. These peers leave the system immediately.

New peers on the other hand will join the `one-club' state with high likelihood, since most candidate peers to download from already belong to the `one-club'. This re-creates the missing chunk syndrome, as each chunk downloaded from the server corresponds to a peer leaving the system, and thus for the system to be stable, we need $\lambda < 1$.

We now prove Theorem~\ref{theo:stable}.

\section{Proof of Theorem~\ref{theo:stable}}
\label{sec:proof}

Assume we have $S$ peers in the system. We denote by $\alpha_i$ the fraction of the $S$ peers that have $i$ chunks. That is, $\alpha_0 S$ peers have no chunks, $\alpha_1 S$ have received one, etc. Denote by $M$ the number of chunks which need to be downloaded to satisfy all the customers in the system. Namely, $M = \Sigma_{i=0}^{k-1} \alpha_i (k-i)$, as $\alpha_0 S$ peers need $k$ distinct chunks to leave the system, $\alpha_1 S$ need $k-1$, etc. We denote by $C_i$ the set of peers with $i$ chunks (as well as its cardinality in a slight abuse of notation).

We will prove that the drift of $M$ is negative for $S$ large enough (since $S \leq M \leq kS$, we can indifferently consider a bound on $S$ or $M$) and then apply Theorem~\ref{theo:drift}).

We now prove a series of Lemmas.

\bLemma
\label{lemma:1}
For $\epsilon > 0$, if there exists $i \geq 0$ such that $\epsilon < \alpha_i < 1-\epsilon$, then
\begin{eqnarray}
\Delta M < \lambda k - \epsilon^2 S
\end{eqnarray}
\eLemma

\bProof
Consider $i \geq 0$ such that $\epsilon < \alpha_i < 1-\epsilon$. Then the peers in $C_i$ can exchange chunks with all the peers not in $C_i$ (either by giving a chunk to the peers in $C_j$ for $j<i$ or by receiving a chunk from $C_j, j>i$). This means that at each time step, the rate of exchange is at least $\alpha_i(1-\alpha_i)S$. By the Lemma's conditions, $\alpha_i (1-\alpha_i) > \epsilon^2 $, therefore the rate of chunk downloads (i.e. the rate of decrease for $M$) is greater than $\epsilon^2 S$.

The rate of increase of $M$ is $k\lambda$, therefore concluding the proof of the Lemma.
\eProof

We introduce some definitions needed for the next Lemmas. 

\bDefinition We call a {\em primary seed} a peer which received a chunk directly from the server. We call a {\em secondary seed} a peer which received a chunk directly from a primary seed. Finally, we call a {\em tertiary seed} a peer which received a chunk directly from a secondary seed.
\eDefinition

\bLemma
\label{lemma:primary}
At any time in the system, there are always $k-1$ primary seeds in the system which received their chunks within the $k$ previous steps.
\eLemma

\bProof
Per our technical assumption, there is always a new primary seed created at each time step. A primary seed will then stay at least another $k-1$ steps in the system, as it receives at most one chunk per step and needs to find another $k-1$ chunks to leave the system. Therefore at any time step, the primary seed created at this step and at the previous $k-2$ steps are in the system. Therefore there are always $k-1$ primary seeds created within the previous $k$ steps present at any time in the system.
\eProof

\bLemma
\label{lemma:tertiary}
If for $\epsilon>0$ and $S >> A$, if there exists $i<k-1$ such that $\alpha_i> 1-\epsilon$, then there are at least $\beta_i (k-1)$ secondary seeds and $\beta_i (k-1)$ tertiary seeds in the system, where $\beta_i = (\alpha_i - (2k+2A)/S)^2$.
\eLemma

\bProof
There are $k-1$ primary seeds per the previous lemma at any time in the system. Therefore, if we denote the current time by 0, there were $k-1$ primary seeds at step -1 and -2. The $k-1$ seeds gave chunks to peers in $C_i$ at rate at least $(\alpha_i-(2A+k)/S)(k-1)$, since there were at least $S-(2A+k)$ peers at step -2 in the system which would be in $C_i$ at step 0 (not counting the $k-1$ secondary seeds, and since less than $A$ peers can arrive per step) and primary seeds can give chunks to any other older peer by the property of the fountain code. Further, since $i<k-1$, giving a chunk to a peer in $C_i$ will move this peer to $C_{i+1}$, therefore the secondary seeds will not leave the system.
Thus there are $(\alpha_i-(2A+k)/S)(k-1)$ secondary seeds at step 0 and -1. The $(\alpha_i-(2A+k)/S)(k-1)$ secondary seeds at step -1 can all give chunks to older peers in the set of peers that will end up in $C_i$ at step 0 (excluding $2k$ primary and secondary seeds), and there are at least a fraction $\alpha_i - (A+2k)/S$ at step -1, therefore there are at least $(\alpha_i-(A+2)/S)(\alpha_i-(2A+k)/S)(k-1)$ at step 0. Since $\beta_i < (\alpha_i-(A+2k)/S)(\alpha_i-(2A+k)/S)<(\alpha_i-(2A+k)/S)$, this completes the proof of the Lemma.
\eProof

Note that for $i=k-1$, the secondary seeds might be leaving the system right away, so we cannot make a similar statement for $i = k-1$.

\bLemma
\label{lemma:Ci}
For $k>2$, there exists $S'>0$ such that for $S>S'$, if there exists $i < k-1$ such that $\alpha_i > 1 - \epsilon$ at time 0, then
\begin{eqnarray}
\Delta M < (k-i)(2-k)
\end{eqnarray}
\eLemma

\bProof
We are satisfying the conditions of Lemma~\ref{lemma:tertiary}, therefore we have in the system $(k-1)$ primary seeds and at least $\beta_i (k-1)$ secondary and tertiary seeds. That means there are $(k-1) (1 + 2\beta_i)$ seeds at any time step with chunks that are less than $k$ steps old. Since there are at most $kA$ chunks in the system that are less than $k$ steps old, and these have exchanged at rate 1 during these $k$ steps, there are at most $k2^k$ peers with chunks less than $k$ steps old. This means that at least $C_i - k2^kA$ peers have chunks that are older than $k$ steps at time 0. We now consider $S'>0$ such that $k2^kA/S' < \epsilon$.

At steps $-(k-i-2),\ldots,-1$, the primary, secondary and tertiary chunks will exchange with the $(\alpha_i-\epsilon)S$ nodes that have chunks older than $k$ steps which will end up in $C_i$ at time 0. These exchanges will lead to peers that will have at most $i+1$ chunks, and therefore will stay in the system at least $k-i-1$ steps. Therefore all these peers will be in the system at time 0.

There are $(k-1)(1+2\beta_i)(k-i-1)(\alpha_i-\epsilon)$ such peers at time 0, plus the $(1+2\beta_i)(k-1)$ primary, secondary and ternary seeds. Therefore, at time 0, $M$ will decrease by $(k-1)(1+2\beta_i)(k-i)(\alpha_i-\epsilon)^2$ chunks. This is equal to $3(k-1)(k-i)(1-\delta)$ where $\delta$ can be chosen arbitrarily close to 0 by selecting the proper $S'$ and $\epsilon$.

Empty peers arrived in the system at rate $(\lambda-1)$. All the peers that have arrived since time $-(i-1)$ have less than $i$ chunks (they can receive at most one per step) and therefore will exchange with peers in $C_i$ at rate $\alpha_i$. Therefore $M$ decreases from these peers at rate $\alpha_i (\lambda-1) i$.

$M$ increases with rate $\lambda k$. Summing all the contributions, we have:
\begin{eqnarray}
\Delta M & = & \lambda k - \alpha_i (\lambda-1) i - 3(k-1)(k-i)(1-\delta) \nonumber\\
& = & \lambda (k-i) + i -3(k-1)(k-i) \nonumber \\
& = & (k-i)(\lambda - 2(k-1)) + i - (k-1)\nonumber \\
& = & (k-i)(\lambda - k) + (k-i)(2-k) + (i-k+1)\nonumber\\
\end{eqnarray}

The Lemma results from the fact that $\lambda < k$ and $i < k-1$.
\eProof

\bLemma \label{lemma:Ck-1} For $i = k-1$, $\alpha_{k-1}>1-\epsilon$ implies $\Delta M < \lambda - k + \epsilon$. \eLemma

\bProof
By Lemma~\ref{lemma:primary}, we have $(k-1)$ primary seeds in the system at time 0. Therefore $\alpha_{k-1} (k-1)$ chunks will be downloaded from those seeds. The server will provide one chunk to a peer at rate 1. Therefore, the rate of chunk download from the primary seeds is $1 + \alpha_{k-1}(k-1) > k-\epsilon$. The primary seeds will upload to peers in $C_{k-1}$ at rate $(\alpha_i-k/S) (k-1)$ since at least $\alpha_i - k/S$ peers in $C_{k-1}$ have innovative chunks. Therefore, there are $2k-1 - o(\epsilon)$ exchange to/from the primary seeds.

At step $-(k-2), -(k-3),\ldots,0$, empty peers arrived at rate $(\lambda-1)$ and these peers are still in the system. Thus they will download chunks with peers in $C_{k-1}$ at rate $\alpha_{k-1}$. Therefore, the rate of chunk exchange from these peers is thus $\alpha_{k-1} (\lambda-1)(k-1)$.

The rate of increase of $M$ is $\lambda k-1$. Putting it all together:
\begin{eqnarray}
\Delta M & = & \lambda k - 1 -  (2k-1) - (\lambda-1)(k-1) +o(\epsilon) \nonumber\\
 & = & \lambda - k + o(\epsilon)
\end{eqnarray}
\eProof

We can now prove our Theorem. Set $\epsilon >0$ small enough and $S'$ large enough so that the drift of $M$ is negative in all the Lemmas. At any point in time, either one of the $\alpha_i$ satisfies $\epsilon < \alpha_i < 1-\epsilon$ (case 1), or there exists $j$ such that $\alpha_j > 1-\epsilon$ and $\alpha_i < \epsilon$ for all $i \neq j$ (case 2).

In case 1, apply Lemma~\ref{lemma:1}; in case 2, if $i<k-1$, apply Lemma~\ref{lemma:Ci}; otherwise, if $i=k-1$, apply Lemma~\ref{lemma:Ck-1}. In all cases, there exist $\delta > 0$ such that $\Delta M < -\delta < 0$ for $S>S'$. Applying Theorem~\ref{theo:drift} concludes the proof of our Theorem.

\section{Practical Considerations}
\label{sec:prac}

\subsection{Important Observations}

The first important observation is as follows: for $i = k-1$, we have seen that the drift of $M$ is bounded by $\lambda - k$ in Lemma~\ref{lemma:Ck-1}. This means that $\lambda < k$ is not only a sufficient condition for stability, but also a necessary condition.

The second important observation is that the parallelism of the system is conditioned by the number of chunks. The more chunks are needed to retrieve the file, the more peers can be served concurrently by the system. This is a qualitatively different stability than the typical system where, due to the "missing chunk" syndrome, the bottleneck is the rate at which the server can deliver chunks.

We state a corollary of Theorem~\ref{theo:stable}.

\begin{corollary}
As the server sees $\lambda$ and selects $k$, it can always pick a value such that the system is stable. This thus demonstrates that any workload can be stabilized by the proper selection of the encoding parameter $k$.
\end{corollary}

This is true of the system that is limited by the "missing chunk" syndrome as well. But the relationship between $k$ and the performance of the system is different.

Consider the basic system (without the fountain encoding and the server-prioritizes-empty-peers policy). Its bottleneck is the rate at which peers can download chunks from the server, namely one arrival per each time it takes to download a chunk from the server ($\lambda < 1$ chunk download time). If the bandwidth from the server to the peer is $B$, and if the chunk size is $F/k$ for a file of size $F$ divided in $k$ chunks, then the server can serve $Ck/F$ chunks per second, and the arrival rate of customers has to be less than $Ck/F$. Increasing $k$ therefore increases the capacity of the system {\em linearly}.

In our scheme, for the same file $F$ divided in $k$ chunks and the same bandwidth $C$, $k$ peers can be served per slot ($\lambda < k$). A slot corresponds to the time to exchange a chunk, therefore a slot lasts $F/(kC)$ unit of times, or conversely, $kC/F$ chunks can be served per unit of time. Therefore, the arrival rate of peers in the system can be $k^2C/F$, which is {\em quadratic} in $k$. Dividing the size of the chunks by two quadruple the capacity of the system (without taking into account any complexity trade-offs). We rephrase this as the following theorem:

\bTheorem Increasing the number of chunks of a given file by a factor $\delta$ increases the stability region by a factor $\delta^2$.
\eTheorem

This has a practical impact. Assuming you have $10^6$ peers arriving to download a file, then you would want to divide the file in $k= 10^3$ chunks so that the value $k^2$ matches the arrival rate into the system. As it turns out, a ninety minute movie is decomposed into 540 chunks of 10s or 2700 chunks of 2s, which are the chunk sizes used for adaptive video streaming mechanisms such as DASH~\cite{Stoekhammer2011,grandl2013,Lederer2012} or Apple HTTP Live Streaming. So our scheme is naturally tuned to support a magnitude of the order of million of users using the current chunk size used in existing video streaming mechanisms.

\subsection{Algorithm Design}

We now design an algorithm to stabilize the system: set $\tau$ to be a timer window. Start at time 0. Set $n=0, a = 0$. While $n \tau \leq t < (n+1) \tau$, count the arrival of new peers. Increment $a$ for each new arrival. When $t = (n+1)\tau$, increment $n$. Estimate the arrival rate as $a/\tau$. If $a/\tau > k-1$, then increase $k$ until $k> a/\tau$. Reset $a$ to 0 and iterate. Upon increasing $k$, the server has to notify the new peers arriving after $(n+1)\tau$ to only communicate in between themselves as the chunks are not compatible with the previous chunks (they correspond to different values of $k$). As for the peers in the system before $(n+1)\tau$, the server should assess whether or not the distribution of chunks would allow them to leave the system. Otherwise, it drains this pool of older peers using the capacity freed with the newest peers by the increase of the $k$ parameter.

In order to achieve a proper code in a practical manner, it is possible to split the file into $k$ chunks and use random linear codes over some Galois field, and generate the coefficient randomly for a linear combination of the $k$ chunks. Then $k$ network coded packets will yield $k$ degrees of freedom with high probability (depending on the size of the Galois field). We can generate $K$ different combinations, as the probability of collision for $k$ combinations will correspond to the probability of drawing $k$ different packets out of $K$ possible packets. The server complexity then reduces to generating those linear combinations. This is the encoding we use in the evaluation section below.

This encoding can be used to vary the parameter $k$ as well. Consider a file split into $2^m$ chunks. Then the network coding coefficients of the linear combinations can be such that sequences of 2 or 4 or 8 chunks are preserved. For instance, chunk 1 and 2 can receive the same coefficients, chunk 3 and 4, etc. This would in effect split the file into $2^{m-1}$ chunks and allow for increasing the number of chunks for the same file to $2^m$ later on, or decreasing it to $2^{m-2}$ by merging 1 through 4, 5 through 8, etc.

Such a scheme would allow for dynamically varying the parameter $k$ without having to segregate peers into before and after the epoch when $k$ has changed. An actual implementation of such a varying rate file distribution system is for future work.

\section{Numerical Evaluation}
\label{sec:num}

We implemented the proposed mechanism in Matlab. Rather than using an ideal fountain code, we chose a large value of $K$, and considered Linear Network Coding with $K >> k$ distinct combinations. We set $K= 10,000$ (which means that each chunk is innovative with probability $1-10^{-4}$ rather than 1 in the ideal encoding case) and let the process evolve from empty initial conditions. We implemented both the {\em baseline mechanism}, the proposed mechanism and two intermediate variations of the baseline mechanism, namely one with the server delivering encoded packets to random peers, and one with the server only providing non-encoded packets to "empty" peers (in our implementation, we do not even select an empty peer, but select the latest peer having joined in the system, which has a high likelihood, but may not be, empty).

We took $k = 5$ for both mechanisms, and varied $\lambda$. We declared the system unstable and stop the simulation if the number of peers in the system grew to over 1,000. We let the system run for 10,000 units or time or until an unstable behavior is detected, whichever came first.

\begin{figure}[!t]
\centering
\includegraphics[width=3.5in]{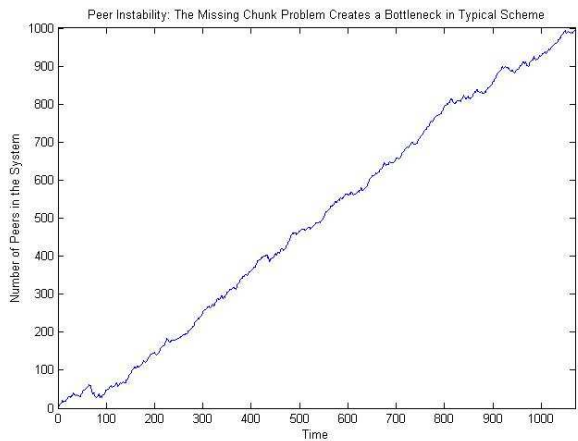}
\caption{Random Peer Contact with No Fountain Code (Baseline Mechanism) $\lambda > 1$}
\label{fig:p2psimbl}
\end{figure}

Figure~\ref{fig:p2psimbl} shows the behavior of the baseline system for $\lambda = 2$ which is well above the server download capacity of 1 of the {\em baseline mechanism}, but well below $k = 5$. As a consequence, the system diverges right away. One can observe that the rate of growth for the number of peers is roughly 1 peer per unit of time, that is $\lambda - 1$, as expected.

\begin{figure}[!t]
\centering
\includegraphics[width=3.5in]{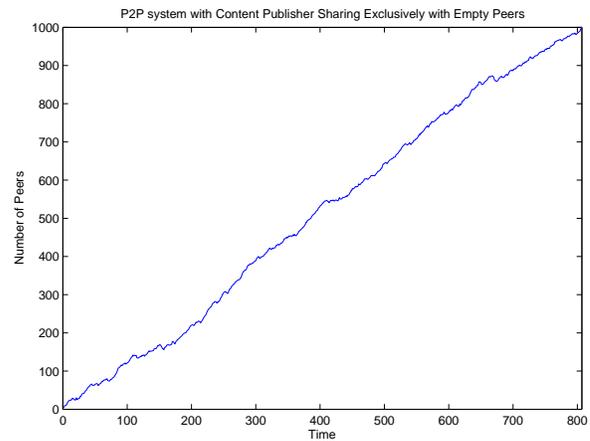}
\caption{Random Peer Contact with Content Server Only Serving Peers in $C_0$, $1 < \lambda < k$}
\label{fig:p2psimlast}
\end{figure}

Figure~\ref{fig:p2psimlast} has the server prioritize the elements of $c_0$ for chunk download. This is one of the mechanisms we use in our proposed P2P protocol. We have discussed why it should not improve the stability in Section~\ref{sec:model}, and the simulations obviously confirm this: it has no consequence on the stability of the system. We do not include a similar plot with only the fountain code implemented (without the prioritization) as it is very similar: using only "baseline + fountain code" does not have an impact on the system performance as well.

\begin{figure}[!t]
\centering
\includegraphics[width=3.5in]{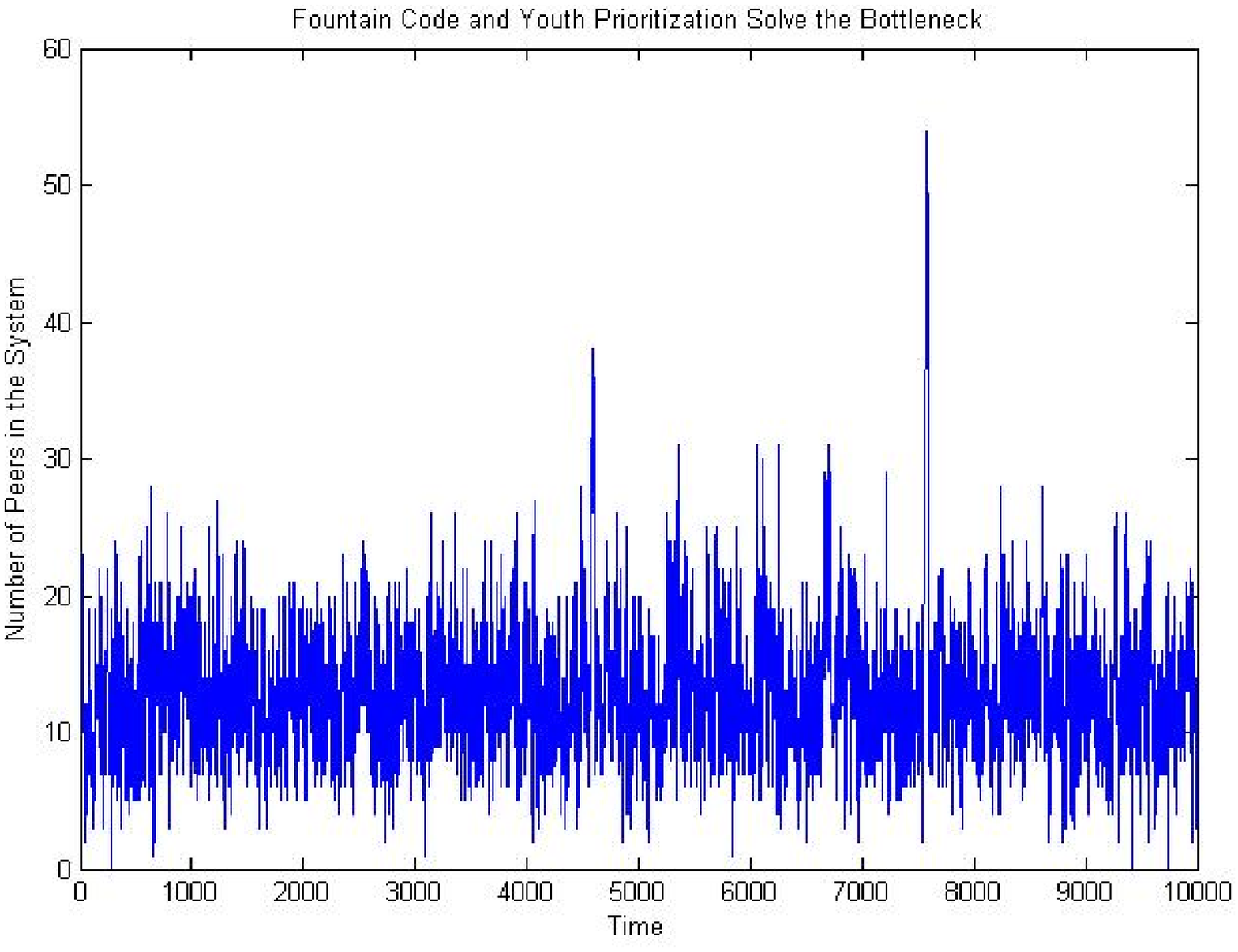}
\caption{Random Peer Contact with Fountain Code and Server-Prioritizes-Empty-Peers policy for $\lambda<k$}
\label{fig:p2psimfc}
\end{figure}

Figure~\ref{fig:p2psimfc} shows now the system corresponding to our proposal under the same assumptions ($\lambda = 2, k = 5$ and server rate equal 1). The number of peers in the system stays roughly constant. It is clear that the system's behavior is dramatically different under this policy as opposed to the baseline policy, as the number of peers in one case grows linearly with time, while in the other, it stays stable for an extended period of time.

\begin{figure}[!t]
\centering
\includegraphics[width=3.5in]{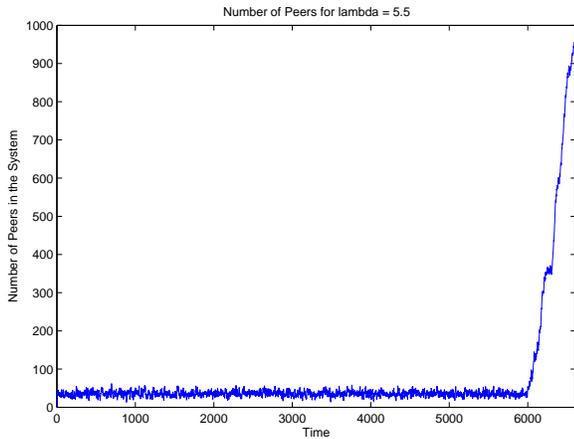}
\caption{Random Peer Contact with Fountain Code and Server-Prioritizes-Empty-Peers policy for $\lambda>k$}
\label{fig:p2psimfc-unstab}
\end{figure}

Note that this is only an indication, not a proof of stability, as it might take a long time for the system to reach a missing chunk syndrome state from its initial conditions. To illustrate this last point, we refer to Figure~\ref{fig:p2psimfc-unstab}, where our proposed mechanism is simulated under $\lambda = 5.5 > 5$. We can see that for a long period of time, it does look stable, until a situation finally arises that the mechanism cannot counteract.

\section{Conclusion}
\label{sec:con}

We have proposed a simple peer-to-peer mechanism which shifts the complexity of stabilizing the system away from the peer-selection and piece-selection mechanism, by using fountain code and a simple prioritization of the content publisher downloads. While this mechanism allows for a greedy behavior from the peers and a random selection of the contact points for download, it improves the stability region of the system, and for proper choices of the system parameters, can arbitrarily handle the peer arrival process.

We have demonstrated the properties of our mechanism through analysis and through some Matlab simulations.

Future work directions include generalizing the system to heterogenous peers with different bandwidth capability and to assess the time spent in the system as compared to other P2P mechanisms.


\section*{Acknowledgment}

The author wishes to extend his thanks to Stratis Ioannidis (Technicolor Labs, Palo Alto) for fruitful conversations on the topic.



%

\bibliographystyle{ieeetr}
\bibliography{P2PContent}

\end{document}